\documentclass[twocolumn]{IEEEtran}
\usepackage[T1]{fontenc}
\usepackage[latin9]{inputenc}
\usepackage{array}
\usepackage{multirow}
\usepackage{amsmath}
\usepackage{amssymb}
\usepackage{graphicx}

\makeatletter

\providecommand{\tabularnewline}{\\}

\title{Analog MIMO RoC Passive Relay for Indoor Deployments of Wireless Networks}

\author{A. Matera\thanks{A. Matera, M. Donati, and U. Spagnolini are with DEIB, Politecnico di Milano: \{andrea.matera,umberto.spagnolini,marcello.donati\}@polimi.it.}, V. Rampa\thanks{V. Rampa (corresponding author) is with IEIIT, National Research Council of Italy (CNR): vittorio.rampa@ieiit.cnr.it}, M. Donati, A. Colamonico\thanks{A. Colamonico is with ABC Progetti, Milano: armando.colamonico@tin.it.}, A. F. Cattoni\thanks{A. Cattoni is with Keysight Technologies: andrea.cattoni@keysight.com.} and U. Spagnolini\thanks{This work has been partially funded by the TRIANGLE project, European Union Horizon 2020 Research and Innovation Programme, Grant No 688712.}
}

\makeatother

\begin{document}
\maketitle

\vskip -0.5cm
\begin{abstract}
Most of the indoor coverage issues arise from network deployments
that are usually planned for outdoor scenarios. Moreover, the ever-growing
number of devices with different Radio Access Technologies (RATs),
expected for new 5G scenarios and to maintain compatibility with older
cellular standards (mostly 3G/4G), worsen this situation thus calling
for novel bandwidth-efficient, low-latency and cost-effective solutions
for indoor coverage. To solve this problem, Centralized Radio Access
Network (C-RAN) architectures have been proposed to provide dense
and controlled coverage inside buildings. However, all-digital C-RAN
solutions are complex and expensive when indoor layout constraints
and device costs are considered. We discuss here an analog C-RAN architecture,
referred to as Analog MIMO Radio-over-Copper (A-MIMO-RoC), that aims
at distributing RF signals indoors over distances in the order of
50 m. The all-analog passive-only design presented here proves the
feasibility of analog relaying of MIMO radio signals over LAN cables
at frequency bandwidth values up to 400 MHz for multi-RAT applications.
After asserting the feasibility of the A-MIMO-RoC platform, we present
some experimental results obtained with the proposed architecture.
These preliminary results show that the A-MIMO-RoC system is a valid
solution towards the design of dedicated 4G/5G indoor wireless systems
for the billions of buildings which nowadays still suffer from severe
indoor coverage issues.

\thispagestyle{empty} \pagestyle{empty}
\end{abstract}

\vskip -0.5cm

\section{Introduction}

\label{sec:Introduction}

5G networks \cite{Agiwal-2016} are already a reality: in 2020, according
to \cite{Commscope-2016}, 96\% of wireless data traffic will originate
or terminate within a building, against 80\% in 2016, with an exponential
increase of indoor wireless market value. However, only few indoor
commercial real estate have dedicated in-building cellular systems
while most indoor areas are served by wireless networks originally
designed and deployed for outdoor. This is the main reason to provide
bandwidth-efficient, ultra-low latency and low-cost solutions to enhance
indoor coverage. Centralized Radio Access Network (C-RAN) enables
such an ambitious goal with full compatibility among different Radio
Access Technologies (RATs) vendors, and still guaranteeing aggregate
service reliability. Even if cell densification is the key enabler
to cope with radio propagation limits, this will introduce unprecedented
complex propagation scenarios, thus calling for a different approach
in the way RAN resources and functions are allocated and handled.
According to this, C-RAN, already deployed in current 4G networks,
will surely play a fundamental role also in 5G networks, even if it
will need to be substantially redesigned to accommodate more demanding
requirements. In 4G networks, Remote Radio Units (RRUs) and BaseBand
Units (BBUs) communicates over a high-capacity link, namely the mobile
FrontHaul (FH), that is usually implemented by fiber optic connections
(Optical Transport Network). This architecture is designed to support
streaming of digitized radio frequency (RF) signals according to specific
digital streaming protocols e.g. the Common Public Radio Interface
(CPRI) \cite{CPRI-2015}. However, it is widely agreed that today
available CPRI-based FH will hardly scale to the increased radio signal
bandwidth foreseen by 5G networks, especially for multiple-antenna
RRUs with massive MIMO constraints. This is due to the fact that Analog-to-Digital
and Digital-to-Analog Converters (ADC/DAC) used in the RRUs, in addition
to uncontrolled end-to-end latency, will also cause a severe bandwidth
explosion, thus surpassing the capacity of current mobile FH links
\cite{Wubben-2014}. To overcome these limitations, new RAN architectures
have been recently proposed with the aim of making the redistribution
of RAN functionalities between BBUs and RRUs more flexible \cite{Bartelt-2015}.
Even the CPRI group has proposed an enhanced version of the standard
that includes intra-PHY functional splits \cite{CPRI-2019}.

\begin{figure*}[t]
\begin{centering}
\includegraphics[clip,scale=0.27]{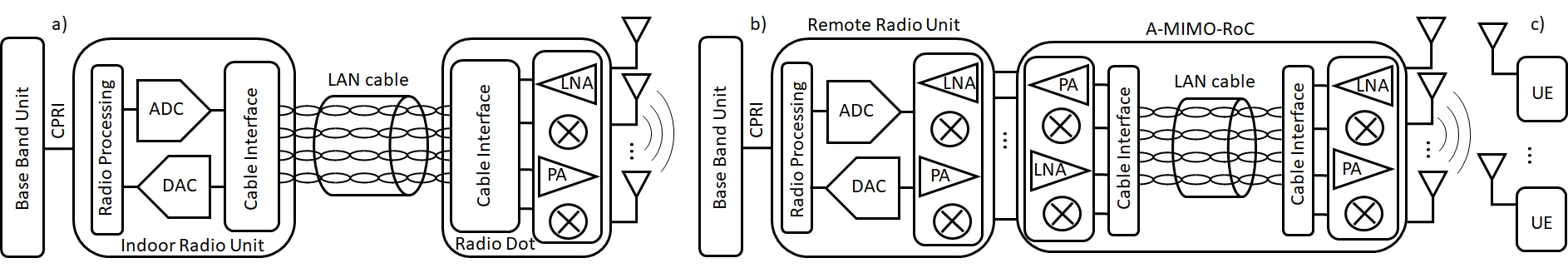}
\par\end{centering}
\caption{\label{fig:1}\protect Architectural comparison: a) Radio Dot connects
the remote antennas with IRU and BBU; b) A-MIMO-RoC connects the remote
antennas with the RRU; c) UEs are connected to a) or b) via the air
interface.}
\vspace{-0.5cm}
\end{figure*}

Unlike digital RAN, Analog Radio-over-Copper (A-RoC) solutions \cite{Gambini-2013,Lu-2014,Tonini-2017}
are able to loosen the FH requirements by simply substituting the
conventional digital FH links with a fully analog transmission of
RF signal between RRUs and BBUs using the ubiquitous cables used for
fixed network deployments. A-RoC FH links not only completely bypass
any bandwidth, latency and complexity issues due to digitization \cite{Acatauassu-2018},
but also reduce hardware costs, improve energy efficiency, and, above
all, easily allow for synchronization among multiple decentralized
RRUs, thus enabling MIMO joint-RRUs processing. In analog C-RAN, the
RF analog-only functionalities are left in the RRUs, which become
based only on low-complexity/low-cost analog components. RRU functionalities
are thus limited to signal conversion to/from RF from/to Intermediate
Frequency (IF) signals, and relaying, adaptation and equalization
of IF signals to/from the BBUs according to the FH capabilities. Furthermore,
the RRUs used in analog FH are protocol-independent and fully-bidirectional
units, hence capable to transparently relay any RAT signal, which
represents a fundamental step towards the wide-range heterogeneity
promised by the 5G revolution. Analog Radio-over-Fiber (A-RoF) \cite{Wake-2010}
is another promising analog FH technology able to provide high throughput
links. However, due to its cost and complexity, it cannot compete
against A-RoC for indoors and it will not be discussed here.

\noindent \textit{Contributions:} The contributions are three-fold:
\textit{i)} the proposal of an all-analog passive-only RF/IF hardware
platform able to transparently transport independent wireless signals,
having different bandwidth requirements, and even belonging to different
RATs, \emph{e.g.} LTE, WiMAX and WiFi, over off-the-shelf Cat-5/6/7
LAN cables; \textit{ii)} the implementation of this platform with
a fully-bidirectional scalable RF/IF architecture with limited cost
and very low complexity; \textit{iii)} the test and verification of
the proposed platform to show that, for limited lengths of the LAN
cables, it can be transparently used by any RF device to connect with
remote RF antennas.

\noindent \textit{Organization:} The A-MIMO-RoC architecture is shown
in Sect.~\ref{sec:Architecture} while its signal model is described
in Sect. \ref{sec:Signal-Model} with some system-level simulations.
Sect.~\ref{sec:Experimental-results} deals with the validation of
proposed architecture and shows some experimental results. Finally,
conclusions are drawn in Sect.~\ref{sec:Conclusions}.

\section{System architecture}

\label{sec:Architecture}

The A-MIMO-RoC architecture presented here finds its roots in the
A-RoC systems proposed by \cite{Gambini-2013} for femto-cell systems
where existing 4 twisted-pairs LAN cables were suggested to decouple
PHY/MAC functionalities from RF units. The A-RoC concept has been
extended to LTE networks in \cite{Medeiros-2016} using 6 twisted-pairs
cables adopted in Digital Subscriber Line scenarios. However, both
proposals restrict the cable bandwidth to a few tenths of MHz. The
Radio Dot System \cite{Lu-2014}, sketched in Fig.~\ref{fig:1}.a,
partly overcame this problem while the A-MIMO-RoC architecture \cite{Matera-2019,Naqvi-2018}
was able to relay multiple RATs MIMO signals over multi-pair copper-cables
by filling the entire cable bandwidth for each pair. The feasibility
of the proposed A-MIMO-RoC system has been shown in \cite{Matera-2019b}
by leveraging the end-to-end testing capabilities of the TRIANGLE
testbed \cite{Zayas-2018}. Fig.~\ref{fig:1}.b shows the generic
block diagram of the A-MIMO-RoC architecture \cite{Matera-2019,Matera-2019b}
where the BBU communicates with the terminals (UEs) as in Fig.~\ref{fig:1}.c,
through the RRU by using remote antennas. Unlike the Radio Dot System
\cite{Lu-2014}, where the RRU is split into two parts, namely the
Radio Dot and the Indoor Radio Unit (IRU), the proposed A-MIMO-RoC
architecture does not have direct access to the BBU control signals.
To limit the cost, simplify the design and increase its applicability,
the A-MIMO-RoC architecture is composed by two identical units called
LAN-to-Coax Converters (L2CCs) that take the RF signals directly from
the RRU (not the BBU) and convert them from RF to IF and vice versa
to connect with the remote RF antennas by using a LAN cable.
\begin{figure}[!b]
\begin{centering}
\includegraphics[clip,scale=0.95]{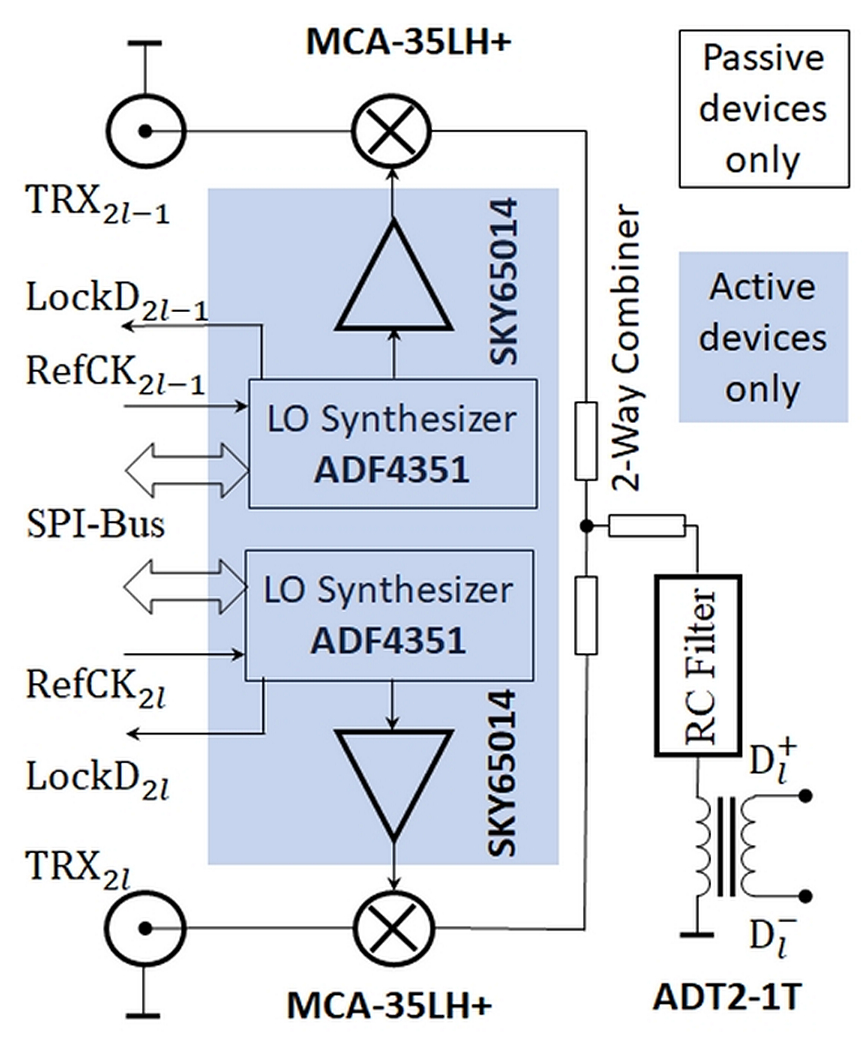}
\par\end{centering}
\caption{\label{fig:2}\protect Detailed functional scheme of the $l$-th
signal slice of the L2CC box shown in Fig.~\ref{fig:1}.b and~\ref{fig:3}.
IC part numbers specify the devices used.}
\end{figure}
\begin{figure*}[t]
\begin{centering}
\includegraphics[clip,scale=0.35]{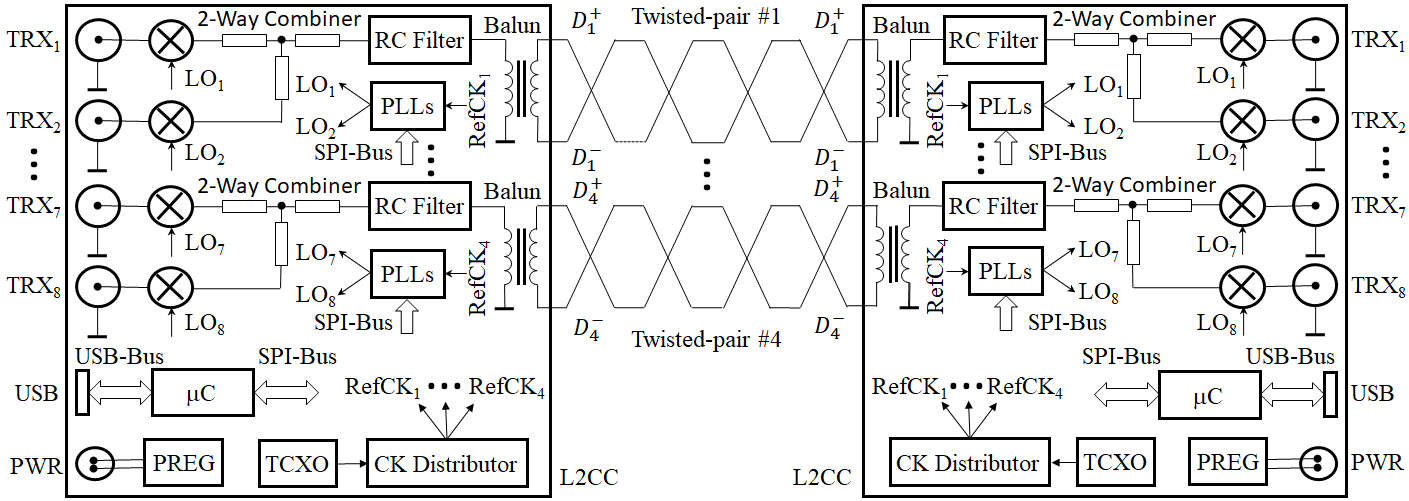}
\par\end{centering}
\caption{\label{fig:3}\protect Block diagram of the all-analog A-MIMO-RoC
system of Fig.\ref{fig:1}.b. The L2CC boxes are composed by: \emph{i})
$L$ signal slices, detailed in Fig.~\ref{fig:2}; \emph{ii}) the
clock generator and distributor; \emph{iii}) the microcontroller ($\textrm{\ensuremath{\mu}C}$)
and \emph{iv}) the power regulators (PREG). In this implementation,
it is $N=8$, $M=2$, and $L=4$.}
\end{figure*}
While in \cite{Lu-2014} the twisted-pairs of the LAN cable carry
also control, signalling, synchronization lines and power, the bi/directional
L2CC blocks of Fig.~\ref{fig:1}.b are used only to adapt, convert
RF, and equalize IF signals. The A-MIMO-RoC prototype is implemented
using only passive devices in the RF/IF signal path with no LNAs (Low
Noise Amplifiers), PAs (Power Amplifiers) or any active filter/mixer
as in the generic architecture of Fig.~\ref{fig:1}.b. Active devices
could be used in the RF/IF paths but this would imply a more complex
and costly implementation. Each L2CC is composed by $L$ identical
signal slices (Fig.~\ref{fig:2} and~\ref{fig:3}) and each $l$-th
twisted-pair is assigned to the $l$-th signal slice with $l=1,...,L$.
Assuming $N=ML$ RF signals (it is $N=8$, $M=2$, $L=4$ in our implementation),
the $l$-th signal slice takes $M$ RF signals, that are applied to
the RF ports TRX$_{M(l-1)+1}$ up to TRX$_{Ml}$. Then \textit{i)}:
each signal slice converts the RF signals into $M$ IF ones by using
the doubled-balanced diode-based mixers MCA-35LH+ from Mini-Circuits.
The Phase-Locked Loop (PLL) synthesizer ADF4351, from Analog Devices,
and the amplifier SKY65014-70LF, from Skyworks Solutions, are used
for each Local Oscillator (LO) stage. The $M$ IF-converted signals
are then: \textit{ii)} combined in a single IF line, using a resistive
M-way combiner, and: \textit{iii)} fed into the passive cable equalizer
(split into two identical RC filters) that is tuned for fixed cable
lengths (Sect.~\ref{sec:Experimental-results}). Finally: \textit{iv)},
the balun ADT2-1T+, from Mini-Circuits, is used to adapt the equalizer
impedance (50 $\Omega$, unbalanced) to the twisted-pair one (100
$\Omega$, balanced). Another L2CC box (Fig.~\ref{fig:3}) is then
employed to convert IF signals back to RF just reversing the previous
operations without using any control signal: Uplink (UL) and Downlink
(DL) modes are thus identical. LO frequencies are statically assigned
during start-up by programming the PLLs using the microcontroller
PIC16F1789 ($\textrm{\ensuremath{\mu}C}$) from Microchip Technology.
LOs can be also dynamically configured through remote commands from
the USB interface by using the internal SPI Bus. The clock generator
feeding the PLLs is composed by the 10 MHz TCXO (Temperature-Compensated
Crystal Oscillator) CW837CT-ND from Connor-Winfield (frequency stability
$\pm$ $50$ ppb), while the clock distributor is the IDT 83908I-02.
To avoid non-linearity effects, the recommended maximum input power
for each RF port is set to 0 dBm while +5 dBm is the maximum applicable
RF input power. The current L2CC implementation shown in Fig.~\ref{fig:3}
is composed by a motherboard that holds the power regulators (PREG),
the $\textrm{\ensuremath{\mu}C}$, the USB interface, the clock generator
(TCXO), clock distributor (CK Distributor), and $L=4$ daughter-boards
that implement the signal slices. Scalability is obtained by \emph{i})
increasing the number $L$ of daughter-boards, if LAN cables with
$L>4$ twisted-pairs are used, by adding more clock lines; \emph{ii})
increasing the number $M>2$ just adding more mixers and combiners.
Theoretically, the proposed platform could be employed in the mm-Wave
bands of the 5G New Radio Frequency Range 2 (FR2) allocation. However,
transmission of wide-band non-OFDM signals is strongly impaired by
the LAN cable characteristics and full active cable equalization must
be implemented (Sect.~\ref{sec:Experimental-results}). Presently,
the high cost of current mm-Wave devices and their limited performances
restrict the applicability of the proposed passive architecture to
the FR1 bands only. Finally, Tab.~I shows the main differences between
available C-RAN architectures where PoE indicates Power over Ethernet
and PO means Propagation Only due to the delay of the analog and/or
optical components. Moreover, unlike A-RoC systems \cite{Lu-2014,Medeiros-2016},
the SF2SF2 feature allows us to use $M$ signals/pair at a time (\emph{e.g.,}
for MIMO) not only $1$ signal/pair.

\begin{table}
\centering
\centering{}\caption{Comparison of main features of various C-RAN systems.}
\begin{tabular}{|c|c|c|c|c|}
\hline 
\multirow{2}{*}{System} & CPRI & A-RoF & A-RoC & A-MIMO-\tabularnewline
 & \cite{CPRI-2015} & \cite{Wake-2010} & \cite{Lu-2014,Medeiros-2016} & RoC \cite{Matera-2019b}\tabularnewline
\hline 
BW increment & $\times18\:\div30$ & $\times1$ & $\times1$ & $\times1$\tabularnewline
\hline 
Latency & $\sim100\,\mu s$ & PO & PO & PO\tabularnewline
\hline 
Complexity & High & High & Low & Very low\tabularnewline
\hline 
RF antennas & $N\gg1$ & $N\gg1$ & $N\leq L$ & $N\leq ML$\tabularnewline
\hline 
Infrastructure cost & Very high & Very high & Very low & Very low\tabularnewline
\hline 
Power supply & External & External & PoE/Ext & PoE/Ext\tabularnewline
\hline 
\end{tabular}
\end{table}

\section{Signal model}

\label{sec:Signal-Model}

\begin{figure}[!b]
\begin{centering}
\includegraphics[clip,scale=0.56]{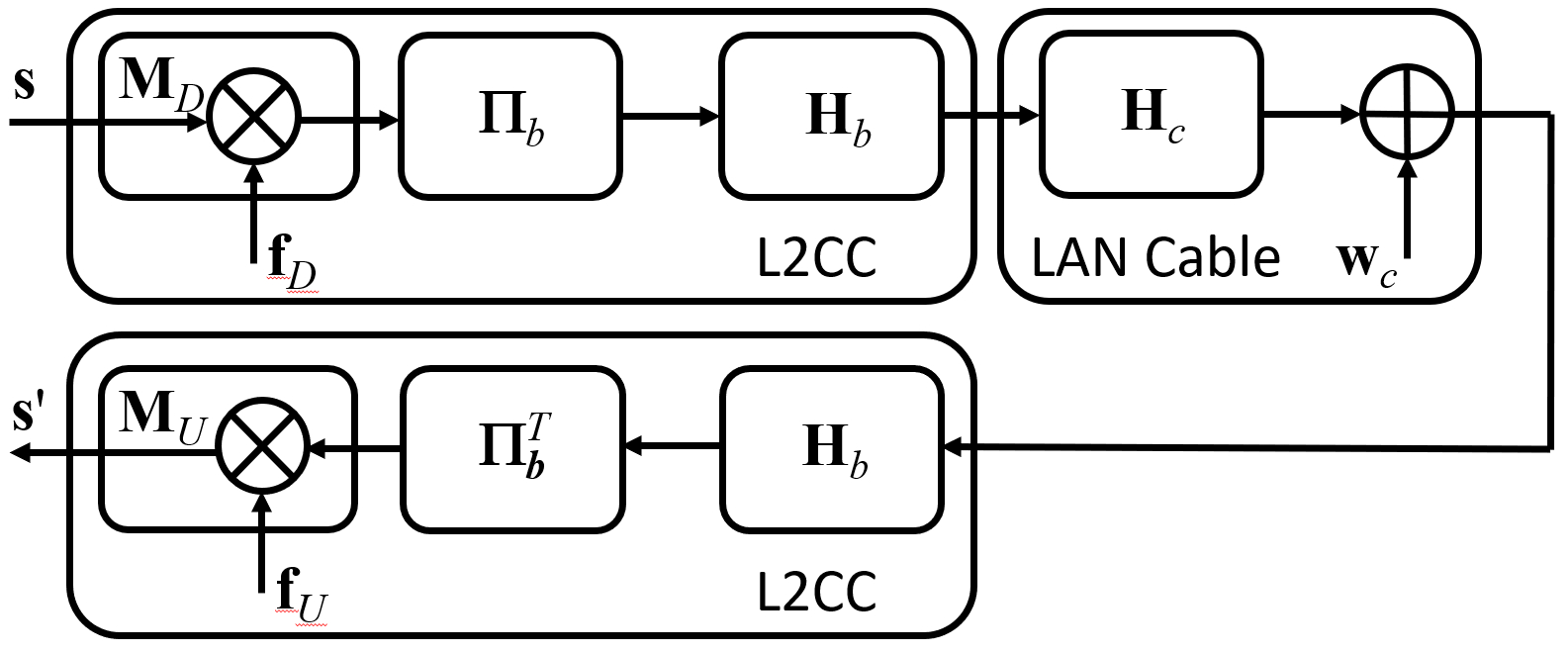}
\par\end{centering}
\caption{\label{fig:4}\protect DL/UL signal model the A-MIMO-RoC platform.}
\end{figure}

As shown in Fig. \ref{fig:4}, the in/out RF signals are represented
by the $N\times1$ band-pass vectors $\mathbf{s}$ and $\mathbf{s'}$
with central frequencies $\mathbf{f}_{0}=[f_{n}^{(0)}]$ and $\mathbf{f'}_{0}=[f'{}_{n}^{(0)}]$,
respectively. Assuming initial zero phase terms, $\left[s_{n}(t)\right]$
and $\left[s'_{n}(t)\right]$ as baseband representations of $\mathbf{s}$
and $\mathbf{s'}$, it is $\mathbf{s}=[\Re\{s_{n}(t)\exp(2\pi jf_{n}^{(0)}t)\}]$
and $\mathbf{s'}=[\Re\{s'_{n}(t)\exp(2\pi jf'{}_{n}^{(0)}t)\}]$ while
the signal model is 
\begin{equation}
\begin{array}{l}
\mathbf{s}'=\mathbf{A}\,\mathbf{s}+\mathbf{B}\,\mathbf{w}_{c}\end{array}\label{eq:1}
\end{equation}
with matrices $\mathbf{A}=\mathbf{M}_{U}\,\boldsymbol{\Pi}_{b}^{T}\,\mathbf{H}\,\boldsymbol{\Pi}_{b}\,\mathbf{M}_{D}$
and $\mathbf{B}=\mathbf{M}_{U}\,\boldsymbol{\Pi}_{b}^{T}\,\mathbf{H}_{b}$
of size $N\times N$ and $N\times L$, respectively.
 Noise effects are represented by the $L\times1$ Gaussian vector
$\mathbf{w}_{c}$ while the $N\times N$ matrices $\mathbf{M}_{D}=\textrm{diag}[\cos(2\pi jf_{n}^{\left(D\right)}t)]$
and $\mathbf{M}_{U}=\textrm{diag}[\cos(2\pi jf_{n}^{\left(U\right)}t)]$
represent down- and up-conversion that are driven by the $N\times1$
LO frequency vectors $\mathbf{f}_{D}=[f_{n}^{(D)}]$ and $\mathbf{f}_{U}=[f_{n}^{(U)}]$,
respectively. The matrix $\mathbf{H}=\mathbf{H}_{b}\circledast\mathbf{H}_{c}\circledast\mathbf{H}_{b}$
of size $L\times L$ includes the filtering effects due to $\mathbf{H}_{c}$,
imputable to each twisted-pair of the LAN cable, and those due to
$\mathbf{H}_{b}$, introduced by the mixer, balun and cable equalizer.
Band inversion is avoided by proper programming the LO frequencies
of the corresponding signal slices. Likewise, also due to the filtering
effects introduced by $\mathbf{H}_{b}$ and $\mathbf{H}_{c}$, IF
images are eliminated by selecting the appropriate LO frequencies.
Theoretically, the corresponding LO signals are set to the same frequency
$\mathbf{f}_{D}=\mathbf{f}_{U}$. In practice, fine LO tuning can
be used to reduce frequency errors between in/out RF signals (Sect.~\ref{sec:Experimental-results}).
To mitigate the LAN cable impairments, a key feature of the A-MIMO-RoC
architecture is the mapping of the RF signals of each RRU antenna
into a combination of twisted-pair/IF allocations, namely Space-Frequency
to Space-Frequency (SF2SF) mapping \cite{Matera-2019}. Here, SF2SF
mapping is split into Space and Frequency Mappings: the $L\times N$
binary matrix $\boldsymbol{\Pi}_{b}=[(\pi_{b})_{l,n}]$ defines the
Space Mapping matrix that implements only the space allocation strategy.
It is statically assigned by selecting the chosen RF cable connections
since the signal slice/twisted-pair association is fixed. In fact,
the RF signals applied to the TRX$_{2l-1}$ and TRX$_{2l}$ ports
are converted, mixed together and then assigned to the $l$-th signal
slice that is wired to the $l$-th twisted-pair: $\forall n=1,...,N$
and $\forall l=1,...,L$, each binary element $(\pi_{b})_{l,n}$ is
set to $1$ iff the $n$-th RF signal is connected to the $l$-th
signal slice, otherwise it is set to $0$. Frequency Mapping is dynamically
performed by selecting the specific LO frequencies $\mathbf{f}_{D}=[f_{n}^{(D)}]$
and $\mathbf{f}_{U}=[f_{n}^{(U)}]$.

\begin{figure}[h]
\begin{centering}
\includegraphics[clip,scale=0.35]{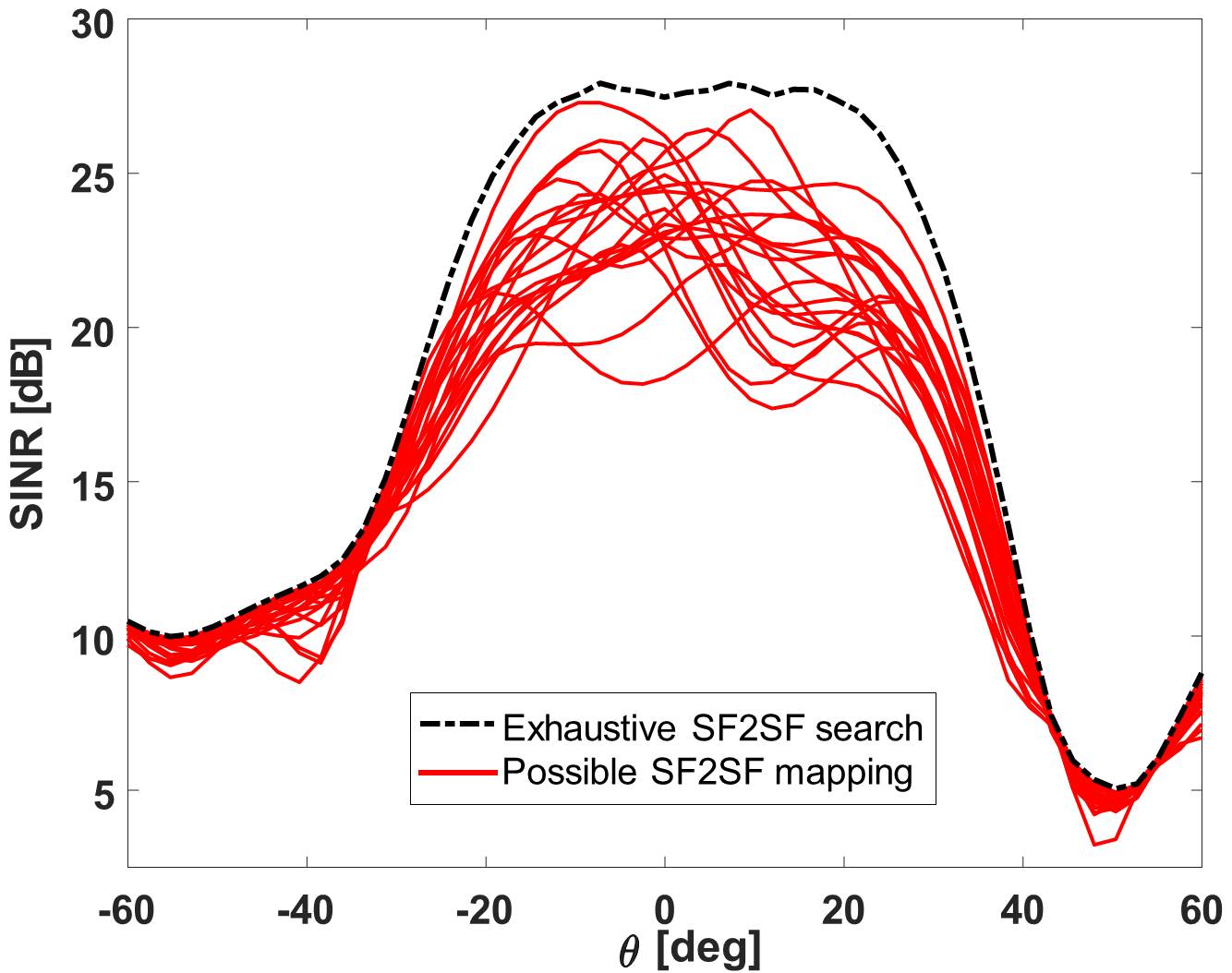}
\par\end{centering}
\caption{\label{fig:5}\protect SINR vs. $\theta$: exhaustive search vs.
some possible SF2SF mappings.}
\end{figure}

Adopting the model (\ref{eq:1}) and the ideal interference scenario
shown in Sect. IV of \cite{Matera-2017} with one active UE, two interfering
UEs, 20 MHz LTE signals with QAM modulation, $L=4$ twisted-pairs
and $N=8$ antennas beamforming at the BBU, Fig.~\ref{fig:5} shows
the SINR (Signal to Interference-plus-Noise Ratio) vs. $\theta$ (UE
steering vector). Uniform linear array and Minimum Variance Distortionless
Response (MVDR) beamforming are also assumed. The optimal SF2SF mapping
is shown with possible specific mapping choices. It is apparent that
the SINR dispersion is $\sim10$ dB from the exhaustive search of
the optimum mapping. For the experimental platform, the optimal choice
of the SF2SF values are set up according to the evaluated end-to-end
performances (Sect.~\ref{sec:Experimental-results}).

\section{Measurements and experimental results}

\label{sec:Experimental-results}

\begin{figure}[tb]
\begin{centering}
\includegraphics[clip,scale=0.93]{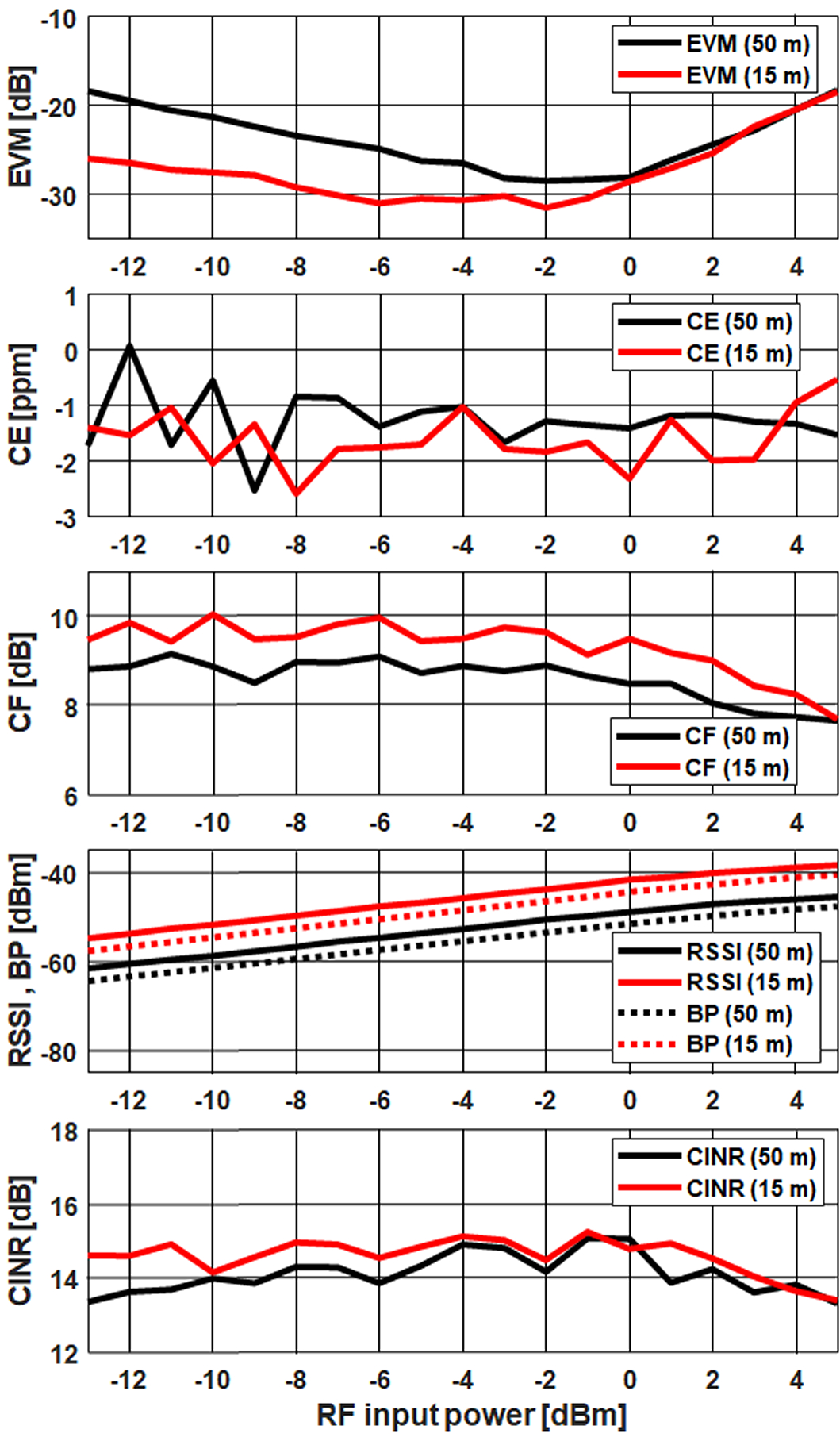}
\par\end{centering}
\caption{\label{fig:6}\protect From the top: EVM {[}dB{]}, CE {[}ppm{]},
CF {[}dB{]}, RSSI {[}dBm{]}, BP {[}dBm{]} and CINR {[}dB{]} measurements
with WIMAX signals. Both results for $15$ m and $50$ m long LAN
cables are considered.}

\vspace{-10pt}
\end{figure}

\begin{figure}[t]
\begin{centering}
\includegraphics[clip,scale=0.57]{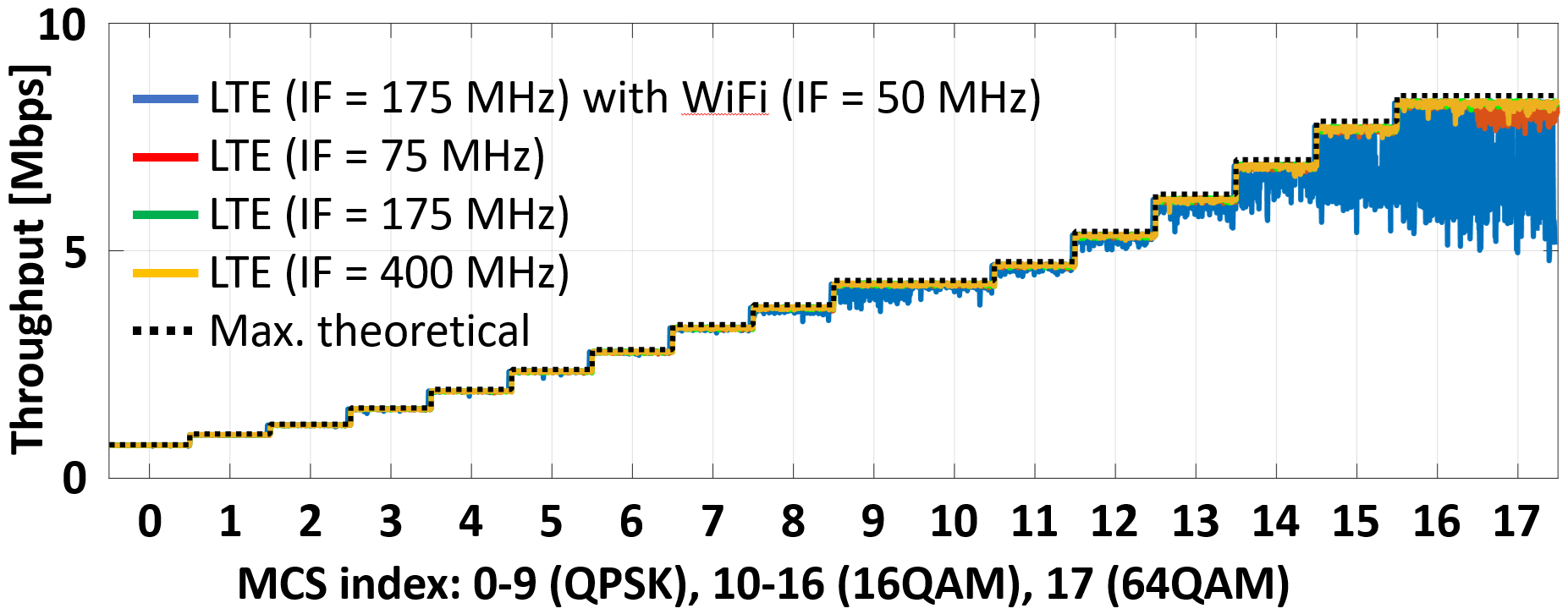}
\par\end{centering}
\caption{\label{fig:7}\protect Throughput {[}Mbps{]} of end-to-end tests
for different protocols (LTE, WiFi), MCS indexes (0-17), and IF values
(50, 75, 175 and 400 MHz).}
\vspace{-5pt}
\end{figure}

The A-MIMO-RoC performances with LAN cables of different type and
length (Cat-5, 15 m and Cat-5e, 50 m) are evaluated here when connected
to lab instruments and real communication devices. For lab measurements,
we used a R\&S Vector Signal generator SMJ100A with WiMAX/LTE options
and a R\&S FSG Spectrum Analyzer with WiMAX option only. Fig.~\ref{fig:6}
shows the following WiMAX test results with $15$ m and $50$ m cables:
Error Vector Magnitude (EVM), Clock Error (CE), Crest Factor (CF),
Received Signal Strength Indicator (RSSI), Burst Power (BP), and Carrier
to Interference-plus-Noise Ratio (CINR). The TRX$_{1}$ port of the
first L2CC is tied to the R\&S SMJ100A generator while the TRX$_{1}$
port of the other L2CC is connected to the R\&S FSG Analyzer.
The WiMAX signal has $f_{1}^{(0)}=2.63$ GHz while both PLL synthesizers
are configured to generate LO$_{1}$ frequencies $f_{1}^{\left(D\right)}=f_{1}^{\left(U\right)}=2.77$
GHz corresponding to an IF value of $140$ MHz. WiMAX bursts are generated
with the following specs: IEEE 802.16-2004 $8$ ms TDD DL burst, guard
time $1/4$, roll-off factor $0.1$, Modulation and Coding Scheme
(MCS) set to 16-QAM rate $3/4$, nominal bandwidth $7$ MHz, $16$
sub-channels, and long preamble. From EVM, BP, CF and CINR measurements,
it is apparent the non-linearity effect, mostly due to the mixers,
that limits the RF input power larger than $\simeq0$ dBm for both
cables. For RF input power values lower than $-$7 dBm ($50$ m) and
$-$15 dBm ($15$ m), the signal to noise ratio decreases thus increasing
the EVM over the WiMAX limit of -25 dBm for the selected MCS configuration.
RSSI measurements highlight the effects due to both LAN cable lengths
and passive devices that introduce $\sim50$ dB of end-to-end attenuation
for the 50 m cable and $\sim42$ dB for the 15 m one. Carrier Frequency
Error (CFE), not shown in Fig.~\ref{fig:6}, is almost constant with
small variations in the range $[-4190,-4177]$ Hz (50 m) and $[-3559,-3475]$
Hz (15 m) due to the tuning mismatch between the corresponding LO
lines $f_{1}^{(D)}$ and $f_{1}^{(U)}$ generated by different PLL
synthesizers. CFE can be almost eliminated by tuning both LO frequencies.
The good frequency stability of the clock generation/distribution
stages is also confirmed by the CE tests whose values are in the range
of $[-2.5,+0.1]$ ppm for both cables. Moreover, phase noise measurements
at the $\textrm{LO}{}_{1}$ lines, not shown here, indicate a residual
PM of $1.21$ deg and a RMS jitter of $1.4$ ps. After LO calibration,
the CFE can be reduced below the $\pm0.1$ ppm LTE limit. For end-to-end
evaluation, LTE signals have been generated by a Keysight UXM RAN
emulator with ETSI Static MIMO channel \cite{Zayas-2018}. Fig.~\ref{fig:7}
shows the throughput results in Mbps, for MIMO LTE tests (TDD mode,
$\textit{BW}=5$ MHz, TM3 mode, B38 band, TX power at -20 dBm, 2x2
MIMO) at different IF frequencies ($75$, $175$, and $400$ MHz)
for different MCS indexes. The 2x2 MIMO LTE signals have been relayed
over two twisted-pairs but at the same cable IF, and thus interfering
with each other. Nevertheless, the performance loss is negligible
for almost all MCS and IF values. Fig.~\ref{fig:7} depicts the throughput
results when a WiFi signal at $\textit{IF}=50$ MHz is added to the
LTE one at $\textit{IF}=175$ MHz. A commercial WiFi access point
has been used to handle the WiFi traffic (IEEE802.11n, $\textit{BW}=20$
MHz, 2x2 MIMO, Channel 1, TX power at 0 dBm) and the first two twisted-pairs
have been used to relay the 2x2 MIMO signals. Up to MCS 13 (16QAM)
the effect of the WiFi signal is almost negligible while it is more
pronounced for higher MCS values. These effects are mostly due to
the fact that the LTE UE operates close to its sensitivity threshold
for the high attenuation introduced by the LAN cable. With shorter
LAN cables with length in the order of $15$ m $\div$ $20$ m and
attenuation about $20$ dB $\div$ $25$ dB, the overall system attenuation
problem is no more an issue (see EVM values in Fig.~\ref{fig:6}).
Even if passive implementation simplifies the design, experimental
results for long cables applications and new scenarios in the 5G mm-Wave
bands suggest the use of active devices in the signal path. Finally,
additional test results are presented in \cite{Matera-2019b}.

\vskip -0.2cm

\section{Conclusions and future works}

\label{sec:Conclusions}

This paper presents the architecture and preliminary results showing
the feasibility of the A-MIMO-RoC system. It is able to transparently
transport multi-RAT MIMO signals over a single LAN cable by a judicious
exploitation of the bandwidth of the interconnecting LAN cable up
to several hundreds of MHz. Performance degradation experienced for
high modulation schemes is mainly due to the high attenuation introduced
by the system caused by the full-passive analog implementation. However,
with shorter LAN cable length the overall system attenuation can be
reduced to a tolerable level.

\vskip -0.5cm

\bibliographystyle{IEEEtran}

\end{document}